\documentclass[
    aip,jmp,
    amsmath,amssymb,
    reprint,onecolumn,
   ]{revtex4-2}

\usepackage[utf8]{inputenc}
\usepackage[T1]{fontenc}
\usepackage{mathptmx}
\usepackage[cal=boondoxo,scr=pxtx]{mathalfa}

\usepackage[colorlinks,hyperindex]{hyperref}  % Si on compile en PdfLaTeX
\hypersetup{
   colorlinks,%
   citecolor=blue,%
   linkcolor=blue,%
   urlcolor=blue,%
 }
\usepackage{orcidlink}

\usepackage{amsthm}
    \theoremstyle{plain}
        \newtheorem{theorem}{Theorem}
        
        \newtheorem{property}{Property}

    \theoremstyle{definition}

    \theoremstyle{remark}
        \newtheorem*{remark}{Remark}

\def\setR{\mathbb{R}}

\def\L{\mathscr{L}}
\DeclareMathOperator{\sgn}{sgn}
\DeclareMathOperator{\Lie}{\mathscr{L}}

\newcommand{\sss}[1]{\scriptscriptstyle #1} 

\DeclareMathOperator{\Lrcorner}{\smash{\rule{5pt}{.5pt}\rule{.5pt}{7pt}}} %pas 
\renewcommand{\leq}{\leqslant}

\begin{document}

\title{Laplacian and codifferential operators on p-forms in (Anti)-de~Sitter spaces: Restriction and continuation}

\author{E.~Huguet \orcidlink{0000-0002-0537-7750}}
    \affiliation{Universit\'e Paris Cité, APC-Astroparticule et Cosmologie (UMR-CNRS 7164), 
    Batiment Condorcet, 10 rue Alice Domon et L\'eonie Duquet, F-75205 Paris Cedex 13, France.}
    \email{huguet@apc.univ-paris7.fr}
\author{J.~Queva \orcidlink{0000-0001-8280-1925}}
    \affiliation{Universit\'e de Corse -- CNRS UMR 6134 SPE, Campus Grimaldi BP 52, 20250 Corte, France.}
    \email{queva@univ-corse.fr}
\author{J.~Renaud \orcidlink{0000-0002-3284-5003}}
    \affiliation{Universit\'e Gustave Eiffel, APC-Astroparticule et Cosmologie (UMR-CNRS 7164), 
    Batiment Condorcet, 10 rue Alice Domon et L\'eonie Duquet, F-75205 Paris Cedex 13, France.} 
    \email{jrenaud@apc.in2p3.fr}

\begin{abstract}
 We derive explicit restriction and continuation formulas between $n$-dimensional (Anti)-de~Sitter spaces and the  $(n+1)$-dimensional Minkowskian ambient space for the codifferential and 
 Laplace-de~Rham operators acting on $p$-forms.
\end{abstract}

\maketitle

\section{Introduction}

The fact that (Anti-)de Sitter [hereafter (A)dS] spaces can be defined as submanifolds of a flat space has long been known, in particular, to describe the symmetry of the space [SO(4,1) or SO(3,2)].
This ambient space approach to (A)dS spaces has been proven useful many times over.\cite{Fronsdal:1974ew, Gazeau:1987nu,Takook:2014paa,Pethybridge:2021rwf}
On another hand, in field theory in curved spaces it is natural to consider $p$-form fields, whether it be in cosmology,\cite{Duncan:1989ug,Koivisto:2009sd} supergravities,\cite{Baulieu:1985md} gauge theories\cite{Henneaux:1986ht} or in other contexts (including their quantization\cite{Folacci:1990eb,Folacci:1990ea}).
Since we can identify $p$-forms on (A)dS with certain $p$-forms on the ambient spaces (namely the transverse and homogeneous forms), it is natural to ask whether the usual calculations ($\delta$ and $\square$ for example) carried out on (A)dS can be replaced by calculations (hopefully simpler) in the ambient space?

In this article, we provide a rigorous answer to this question in the general framework of $p$-forms on  $n$ dimensional spaces (Theorem~\ref{Th-Continuation}).
We start by addressing the problem of restriction in this same general framework (Theorem~\ref{Th-Restriction}).
Results for the Laplace-de Rham operator are  translated  to the Laplace-Beltrami operator  thanks to the Weitzenböck formula (Theorems~\ref{Th-Restriction-LB} and \ref{Th-Continuation-LB}).
The Euclidean case is also considered. 
Specifically,
we relate  the Laplace-de Rham, and thus the codifferential, operator acting on 
$p$-forms  defined on $n$ dimensional (A)dS spaces to their counterparts defined on ambient $n+1$ space.
We derive expressions which relate explicitly differential operators, acting on $p$-forms, on (A)dS and ambient spaces. 
These relations work in both directions, restriction and continuation.

The present work, thus adds tools to the ambient space formalism to handle $p$-forms on (A)dS space.
This, compared to previous endeavors,\cite{Fronsdal:1974ew, Gazeau:1987nu,Takook:2014paa,Pethybridge:2021rwf} without relying on a specific choice of coordinates. 
This, in particular,  simplifies and extends greatly our previous results, \cite{Huguet:2022rxi} where only one-forms  and restriction were considered. Moreover, it sets a firmer ground to generalize these results to Friedmann-Lemaître-Robertson-Walker (FLRW) spaces. 
As it encompasses it, the reader is referred to Ref.~\onlinecite{Huguet:2022rxi} for detailed references and discussions, as they will not be reproduced here.

Our notations and the geometric framework are settled Sec.~\ref{SEC-DefinitionNotations}.
The main results appear Sec.~\ref{SECRestrictionAndContinuation}, where restriction and continuation Theorems are given.
Possible generalizations to FLRW spaces and concluding remarks appear Sec.~\ref{SEC-Conclusion}.
A concise overview of differential geometric tools used in this paper is provided App.~\ref{APP-GeoDiff}.
Necessary properties concerning operations on $p$-forms basis are collected together with their proofs App.~\ref{APPENDIX-Properties-of-Basis}.
Detailed proofs of the main theorems are gathered App.~\ref{SEC-Proofs}. 

\section{Definitions and Notations}\label{SEC-DefinitionNotations}

\subsection{Geometric context}

Our conventions on indexes read: $A,B,C,\ldots=0,\ldots ,n$ and $\lambda,\mu,\nu,\ldots =0,\ldots, n-1$. The cartesian coordinates of the oriented pseudo-euclidean ambiant space $\setR^{n+1}$ are denoted by $\{y^{\sss A}\}$ and the corresponding  coordinate basis $\{\partial_{\sss A}\}$. 
The  ambient space metric is $\eta = \mathrm{diag}(+,-,\ldots,-,-\epsilon)$,   with $\epsilon = 1$ for de~Sitter space and $\epsilon = -1$ for the Anti-de~Sitter space.

The (Anti-)de~Sitter spaces are defined as submanifolds of $\setR^{n+1}$ through $y^2=y_{\sss A}y^{\sss A} = -\epsilon H^{-2}$. Throughout the paper, they are both denoted by $\Sigma$.
For later convenience, let us define the dilation vector $D := y^{\sss A}\partial_{\sss A}$, and 
$h:=(\sqrt{|y^2|})^{-1}$. 
We suppose that $\Sigma$ is oriented by means of its (normalized) exterior normal vector $e_n=hD$. 
Note that $e^n=d(h^{-1})$, so $de^n=0$.

\subsection{Orthonormal basis}\label{SUBSEC-OrthonormalBasis}
Orthonormal frames enjoy simplifying properties regarding the action of various
differential operators. 
We will take full advantage of these properties, in the ambient space context, building convenient orthonormal frames of the ambient space, from that of $\Sigma$. 

To this end, let us consider, in some open set $U\subset\Sigma$, a direct orthonormal 
frame field $\{e_\mu\}$.
Completing this set, with the unit normal $e_n$ at each point of  $U$, leads to a direct orthonormal frame field of $\setR^{n+1}$ defined in each point of $U\subset\Sigma$.
Now, we continue this local orthonormal frame to $\setR^{n+1}$ by homogeneity.
Precisely
at each point $y$ of $\Sigma$ one has $e_{\sss A} = e_{\sss A}^{\sss B} \partial_{\sss B}$, and we can extend $e_{\sss A}^{\sss B}(y)$ to the ambient space by imposing that it be homogeneous of degree zero. This allows 
 us to extend the frame to $\setR^{n+1}$ through the same formula $e_{\sss A} = e_{\sss A}^{\sss B} \partial_{\sss B}$ and the extended vectors $e_{\sss A}$ are now homogeneous of degree $-1$.
Pictorially we just ``pushed'' the frame  along the half-lines emanating from the origin.
Thanks to the homogeneity, the holonomic coefficients of this frame fullfill the very useful relations
\begin{equation*}
    c^n_{\sss AB}=0 \mbox{ and } c^\mu_{\nu n}=h\delta^\mu_\nu.
\end{equation*}
For more details regarding the geometric context and orthornormal basis see App.~\ref{APP-GeoDiff} and Ref.~\onlinecite{Huguet:2022rxi}.

\subsection{Transverse-longitudinal splitting of $p$-forms}\label{SEC-Def-of-splitting}

The identity $j^ni_n + i_nj^n = 1$, in which $i_n$ and $j^n$ stand for $i_{e_n}$ and $j^{e^n}$  (see App.~\ref{APP-GeoDiff} for the notations),
allows us to define transverse and longitudinal parts of
a form: let $\alpha$ be a $a$-form on $\setR^{n+1}$, the adapted splitting in transverse ($\alpha_\perp$), and longitudinal ($ \alpha_\parallel $) parts of $\alpha$, with respect to $\Sigma$, is defined through
\begin{equation}\label{EQ-definition-AdaptedSpliting}
    \alpha := \alpha_\parallel + \alpha_\perp,
\end{equation}
with
\begin{align*}
    \alpha_\parallel 
    & := j_ni^n \alpha 
    = (-1)^{a - 1} (i_n\alpha)\wedge e^n=\frac{1}{(a-1)!}\alpha_{\mu_1\ldots \mu_{a-1} n}
        e^{\mu_1}\wedge\cdots\wedge e^{\mu_{a-1}}\wedge e^{n},\\
    \alpha_\perp 
    & := i^nj_n \alpha 
    = \frac{1}{a!}\alpha_{ \lambda_1\ldots \lambda_a}e^{\lambda_1}\wedge\cdots\wedge e^{\lambda_a}.
\end{align*}
The set of transverse (respectively longitudinal) $p$-forms of $\setR^{n+1}$ will be denoted by \smash{$\Omega_\perp^p(\setR^{n+1})$} [respectively \smash{$\Omega_\parallel^p(\setR^{n+1})$}]. 
A $p$-form, homogeneous of degree $s$, will be termed $s$-homogeneous $p$-form, whose set will be denoted $\Omega_s^p(\setR^{n+1})$.
Finally, the set of $p$-forms, both transverse and $s$-homogeneous, will be denoted by $\Omega_{s \perp}^p(\setR^{n+1})$. 

Let us denote by $m$ the canonical injection map from $\Sigma$ in $\setR^{n+1}$.
Note that, firstly, a $p$-form  $\alpha \in \Omega^{p}(\setR^{n+1})$ is mapped by $m^*$ on its transverse part on $\Sigma$: $\alpha_{\sss \Sigma} := m^* \alpha = \alpha_\perp|_{\sss \Sigma}$, and secondly, a transverse $p$-form  $\alpha \in \Omega^{p}_{\perp}(\setR^{n+1})$, satisfy  $i_n\alpha=0$ and, can be identified  at each point of  $\Sigma$  with  a $p$-form on
$\Sigma$.
This implies: $\forall y_{\sss \Sigma}\in\Sigma,\  \alpha(y_{\sss \Sigma}) = m^*\alpha (y_{\sss \Sigma})$. 
Now, since each point of $\Sigma$ belongs to a half-line emanating from the origin of $\setR^{n+1}$, one can also obtain $p$-forms on $\setR^{n+1}$ from  $p$-forms on
$\Sigma$ thanks to  homogeneous continuation.
Indeed, the pullback $m^*$ induces an isomorphism  between transverse $s$-homogeneous $p$-forms on $\setR^{n+1}$ and those on $\Sigma$:
\begin{equation*}
    \Omega_{s \perp}^p(\setR^{n+1})  \simeq \Omega^p(\Sigma).
\end{equation*}
This allows us to identify $p$-forms on $\Sigma$ with $s$-homogeneous $p$-forms on $\setR^{n+1}$.
Using this identification one can write, in particular, $m^*e_\mu=e_\mu$ while $m^*e_n=0$.
Note also that $m^*h=H$.

\section{Restriction and Continuation of operators}\label{SECRestrictionAndContinuation}

\subsection{The main theorems}

The restriction to $\Sigma$ of a form on $\setR^{n+1}$ is given by the following theorem in which $\square := -(d\delta + \delta d)$ is the Laplace-de Rham operator.
\begin{theorem}[Restriction]\label{Th-Restriction}
Let $\alpha \in \Omega^a(\setR^{n+1})$, $m$ be the canonical injection: $\Sigma \xrightarrow{m} \setR^{n+1}$, and $\alpha_{\sss \Sigma} := m^* \alpha = \alpha_\perp|_{\sss \Sigma}$. Then,
\begin{align}
    \label{EQ-Final-Restrict-delta}
    & m^*\delta_{n+1}\alpha
    = \delta_{\sss \Sigma} \alpha_{\sss \Sigma} 
    - \eta^{nn}m^*[\L_{n}i_{n}\alpha
    + H(n-2a+2)i_n\alpha],\\
    \label{EQ-Final-Restrict-Box}
    & m^*\square_{n+1}\alpha
    = \square_{\sss \Sigma}\alpha_{\sss \Sigma}
    +\eta^{nn} m^*[\L_n^2\alpha+ H(n-2a)\L_n\alpha+2Hdi_n\alpha],
\end{align}
with $\Lie_n = \Lie_{e_n}$ the Lie derivative along $e_n$.
\end{theorem}

On the other hand our considerations on the transverse forms allow us to establish the following converse result:
\begin{theorem}[Continuation]
\label{Th-Continuation}
Let $m$ the canonical injection: $ \Sigma \xrightarrow{m} \setR^{n+1}$, and $\beta_s$ be the $s$-homogeneous (necessarily transverse) extension of $\beta \in \Omega^b( \Sigma)$, that is $\beta := m^*\beta_s$ 
and $\beta_s \in \Omega_{s \perp}^b$. Then, 
\begin{align}
    \label{EQ-Final-Continuat-delta}
    & \delta_{\sss \Sigma}\beta = \delta_{n+1}\beta_s, \\
    \label{EQ-Final-Continuat-Box}
    & \square_{\sss \Sigma}\beta
    = \square_{n+1}\beta_s
    -\eta^{nn}H^2s(s+n-1-2b)\beta_s 
    + 2He^n\wedge(\delta_{n+1}\beta_s).
\end{align}
\end{theorem}
This last theorem calls for some explanations. 
As a consequence of theorem~\ref{Th-Restriction}, taking into account the fact that $\beta$ is transverse, we can prove that
\begin{align*}
    & \delta_{\sss \Sigma}\beta = m^*(\delta_{n+1}\beta_s), \\
    \label{EQ-Final-Continuat-Box}
    & \square_{\sss \Sigma}\beta
    = m^*(\square_{n+1}\beta_s
    -\eta^{nn}H^2s(s + n-1-2b)\beta_s 
    + 2H e^n\wedge(\delta_{n+1}\beta_s)).
\end{align*}
Moreover both arguments of $m^*$, in the above equations, 
are found to be transverse and $(s-2)$-homogeneous.
As a consequence, we can identify $m^*\delta_{n+1}\beta_s$ and $\delta_{n+1}\beta_s$, this is what was done in the first statement.
The same goes for the second statement.

We can see that the calculation of the codifferential on $\Sigma$ is equivalent to that, {\it a priori} simpler, on ambient space.
Now, due to the fact that $\beta$ being transverse does not imply that $d\beta$ is also transverse, the situation is less trivial with regard to the operator $\square$.
Nevertheless it remains quite simple, particularly if we choose $s=0$.

As an illustration, consider a scalar field $\phi$ ($a=0$) and a  $1$-form field $A$ ($a=1$) on (A)dS.
We identify these fields with their homogeneous transverse extensions of degree 0, and consistently keep the same notation for both (A)dS and ambient-space 
fields ($\phi$ and $A=A_\alpha dy^\alpha$).
We then have for all $y\in\Sigma$ 
    \begin{align*}
    &\square_{\sss \Sigma}\phi= \partial^\alpha\partial_\alpha\phi,\\
    &\square_{\sss \Sigma}A=\{\partial^\alpha\partial_\alpha A_\beta+2H^2(\partial^\alpha A_\alpha)y_\beta\}dy^\beta,
    \end{align*}
where $\phi$ and $A$ fulfill $y^\alpha\partial_\alpha\phi=0$ and $y^\alpha A_\alpha=0$ and $y^\beta\partial_\beta A_\alpha=-A_\alpha$.

Proofs of theorems \ref{Th-Restriction} and \ref{Th-Continuation} are given in App.~\ref{SEC-Proofs}. 
These proofs exploit both the simplifications in calculations induced by the orthonormal basis of $\Sigma$ extended to  $\setR^{n+1}$ and the corresponding  splitting of $p$-forms in transverse and longitudinal parts. 
The main part of the proof consists in proving Eq.~\eqref{EQ-Final-Restrict-delta}, the other results are easily deduced from it.

\subsection{The Laplace--Beltrami operator}
Our calculations are based on the  Laplace-de~Rham operator which is well suited to $p$-forms. Nevertheless these results can be transcribed for the Laplace--Beltrami operator since the two are linked by the Weitzenböck formula  (see App.~\ref{APP-GeoDiff}) which, for $\alpha$ a $a$-form, reads
 \begin{equation}\label{EQ-Weitzenbock}
    \square\alpha=\Delta\alpha
    +j^ai^bR(e_a,e_b)\alpha,
\end{equation}
where $R(u,v)=\nabla_u\nabla_v-\nabla_v\nabla_u-\nabla_{[u,v]}$ is the curvature operator, $\nabla$ the Levi--Civita  connection and
${\displaystyle \Delta = \mbox{tr}\nabla^2=g^{ab}(\nabla_a\nabla_b-\nabla_{\nabla_ae_b})}$ is the Laplace--Beltrami operator.

In the case of (A)dS spaces the curvature operator reduces to
\begin{equation}
   R_{\sss \Sigma}(e_\mu,e_\nu)e^\xi
   = \eta^{nn}H^2(\eta_{\mu\lambda}\delta^\xi_\nu
   -\eta_{\nu\lambda}\delta^\xi_\mu)e^\lambda,
\end{equation}
and the Weitzenböck formula simplifies into
\begin{equation}
    (\square_\Sigma-\Delta_\Sigma)\alpha
    =\eta^{nn}H^2a(a-n)\alpha.
\end{equation}
This allows us to rewrite the two previous theorems for the Laplace--Beltrami operators.
\begin{theorem}[Restriction for Laplace--Beltrami operator]\label{Th-Restriction-LB}
Let $\alpha \in \Omega^a(\setR^{n+1})$, $m$ be the canonical injection: $\Sigma\xrightarrow{m} \setR^{n+1}$, and $\alpha_{\sss \Sigma} := m^* \alpha = \alpha_\perp|_{\sss \Sigma}$. Then,
\begin{equation}
    m^*\Delta_{n+1}\alpha
    = \Delta_{\sss \Sigma}\alpha_{\sss \Sigma}
    +\eta^{nn} m^*[\L_n^2\alpha+ H(n-2a)\L_n\alpha+2Hdi_n\alpha+H^2a(a-n)\alpha].
\end{equation}
\end{theorem}

\begin{theorem}[Continuation for Laplace--Beltrami operator]
\label{Th-Continuation-LB}
Let $m$ the canonical injection: $\Sigma\xrightarrow{m} \setR^{n+1}$, and $\beta_s$ be the $s$-homogeneous (necessarily transverse) extension of $\beta \in \Omega^b( \Sigma)$, that is $\beta := m^*\beta_s$ 
and $\beta_s \in \Omega_{s \perp}^b $. Then, 
\begin{equation}
     \Delta_{\sss \Sigma}\beta
    = \Delta_{n+1}\beta_s
    -\eta^{nn}H^2[s^2+ s(n-1-2b) + b(b-n)]\beta_s 
    + 2He^n\wedge(\delta_{n+1}\beta_s).
\end{equation}
\end{theorem}

\subsection{The Euclidean case}

Finally, let us remark that the above theorems transpose to the Euclidean situation  where the embedding of (Anti)-de~Sitter in  the pseudo-Euclidean ambient space  $\setR^{n+1}$  is replaced by that of the sphere $S_n$ in the Euclidean ambient space  $\setR^{n+1}$, the vector $e_n$ becoming the outer normal to the sphere, $\eta^{nn}=1$, and $H^{-1}$ being the radius of the sphere.

\section{conclusion}\label{SEC-Conclusion}

The present work not only extends (to $p$-forms) Ref.~\onlinecite{Huguet:2022rxi} but also introduces simplified calculations and addresses in a detailed way the continuation of operators to ambient space. This, with the help of the embedding in $\setR^{n+2}$ exposed in Ref.~\onlinecite{Huguet:2023xmn}, is a starting point for our ongoing work on FLRW spaces. 
Indeed, in Ref.~\onlinecite{Huguet:2023xmn} we showed that FLRW spaces are obtained through a dilation operator from (A)dS spaces.
We will use this dilation operator to carry differential operators from (A)dS to FLRW in order to obtain, for FLRW in $\setR^{n+2}$, results similar to those obtained in the present paper.
Precisely, we will establish relations between differential operators acting on $p$-forms on a FLRW space and their extensions to ambient space,
where calculations are hopefully simpler.

\appendix

\section{Overview of the main notions of differential geometry used in this work}\label{APP-GeoDiff}

We gather here
some definitions and formulas from differential geometry frequently used in this paper (the notations are those of Ref.~\onlinecite{Fecko}).
We also relate, using our notations, the Laplace-de Rham operator to the Laplace--Beltrami operator through the 
Weitzenböck formula.

Let $(M,g)$ be a smooth pseudo-riemannian manifold of dimension $n$,
$TM$ and $T^*M$ being its tangent and cotangent fiber bundles, respectively.
For $p$ an integer such that $1\leq p\leq n$ then $\Omega^p(M)\equiv \Omega^p = \bigwedge^p T^*M$ is the space of $p$-forms, smooth functions on $M$ are identified with zero-forms: $\Omega^0(M) = C^{\infty}(M,\setR)$.

The natural pairing between a $1$-form $\lambda$ and a vector $v$ is written $\langle\lambda,v\rangle$, the musical applications $\flat$ and $\sharp$ relate this pairing to the metric as $\langle\lambda,v\rangle = g(\sharp\lambda,v)$ and 
$g(u,v) = \langle\flat u,v\rangle.$
We recall that the ``creator operator'' $j_v$ is such that $j_v\alpha = (\flat v)\wedge\alpha$ with $v\in TM$ and $\alpha\in\Omega^a$, with $j_uj_v = -j_vj_u$, and that the interior product  $i_v$  fulfills an antisymmetry relation $i_ui_v = -i_vi_u$, both operators interacting such that
\begin{equation*}
    i_u j_v + j_vi_u = g(u,v),\quad u,v\in TM.
\end{equation*}
The interior product appears in the Lie derivative through Cartan formula: $\Lie_v = i_v d + di_v$.

On an oriented manifold with the Hodge operator $*$, fulfilling $*1=\omega$ and $*\omega = \mathrm{sgn}(g)$ with $\omega$ the volume form,
and for $\alpha\in\Omega^a$ the codifferential is defined as $\delta\alpha = (-1)^a*^{-1}d*\alpha$.
The Hodge operator $*$ interacts with $i_v$ and $j_v$ as
\begin{align*}
    &*i_v\alpha = -(-1)^aj_v*\alpha,\\ 
    &*j_v\alpha = (-1)^ai_v*\alpha,
\end{align*}
with $\alpha\in\Omega^a$ a $a$-form.
See, say, App.~A of Ref.~\onlinecite{Huguet:2022rxi} for additional identities.
These products $i_v$ and  $j_v$ are extended to $1$-forms in a natural way.
For $\lambda\in\Omega^1$ and $\beta\in\Omega^b$ we note
\begin{align*}
    &i^\lambda\beta = i_{\sharp\lambda}\beta,\\
    &j^\lambda\beta = j_{\sharp\lambda}\beta = \lambda\wedge\beta.
\end{align*}
Sometimes we note also the interior product with a $1$-form as 
\begin{equation*}
    \lambda\Lrcorner \beta 
    = i^\lambda\beta = i_{\sharp\lambda}\beta.
\end{equation*}
In an orthonormal basis $(e^a)$, as a short form, we set
    $i^{e^a} = i^a$, 
    $j^{e^a} = j^a$,
from which we obtain the useful formulas
\begin{align*}
    &\ast(e^{a_1}\wedge\cdots\wedge e^{a_p})= i^{a_p}\ldots i^{a_1} \omega,\\
    &\ast i^{a_1}\ldots i^{a_p}\omega=\sgn(g)(-1)^{p(d+1)}e^{a_p}\wedge\cdots\wedge e^{a_1}.
\end{align*}

We can now define the Laplace-de~Rham operator $\Box$, through
\[\Box=-(d\delta+\delta d).\]
This operator is related to the usual Laplace--Beltrami operator.
Indeed, if $\nabla$ is the Levi-Civita connection associated to $g$ then the second covariant derivative 
$\nabla^2_{uv}=\nabla_u\nabla_v-\nabla_{\nabla_uv}$, where $u$ and $v$ are tangent vectors,  brings about the Laplace--Beltrami operator through its trace, namely:
\[ \Delta = \mbox{tr}\nabla^2=g^{ab}(\nabla_a\nabla_b-\nabla_{\nabla_ae_b}).\]
The link between the Laplace-de Rham $\square$ and the Laplace--Beltrami $\Delta$ operators on $p$-forms is given by
 the Weitzenböck formula: 
\begin{equation*}\label{App-EQ-Weitzenbock}
    \square\alpha=\Delta\alpha
    +j^ai^bR(e_a,e_b)\alpha,
\end{equation*}
where $\alpha$ is a $p$-form and $e_a$ any frame,  $R(u,v)=\nabla_u\nabla_v-\nabla_v\nabla_u-\nabla_{[u,v]}$ is the curvature operator and $\nabla$ the Levi-Civita  connection
(see App.~A of Ref.~\onlinecite{Huguet:2022rxi} for a short proof).
We also recall, in any frame $e_a$, the link between the operators $d$ and $\delta$ with the Levi-Civita connection $\nabla$:
$\delta=-i^a\nabla_a$,
$d=j^a\nabla_a$.

\section{Properties fulfilled in orthonormal basis}
\label{APPENDIX-Properties-of-Basis}

We gather here specific properties of orthonormal basis, central for our derivations. 
As explained above (Sec.~\ref{SEC-Def-of-splitting}), we frequently identify $m^*\tau$ with $\tau$ on $\Sigma$ whenever $\tau$ is a transverse $t$-form.

\begin{property}\label{Prop1}
\begin{align*}
    &d(e^{\mu_1}\wedge \cdots\wedge e^{\mu_t})
    = d_{\sss \Sigma}(e^{\mu_1}\wedge \cdots\wedge e^{\mu_t})
    + (-1)^{t} ht\ e^{\mu_1}\wedge  \cdots\wedge e^{\mu_t}\wedge e^n,\\
    &d(e^{\mu_1}\wedge \cdots\wedge e^{\mu_t}\wedge e^n)
    = (d_{\sss \Sigma}e^{\mu_1}\wedge \cdots\wedge e^{\mu_t})\wedge e^n.
\end{align*}
\end{property}
\begin{proof}
Noting that: $de^{n}=0$ and
\begin{align*}
    de^\mu
    =-\frac{1}{2}c^\mu_{\sss AB}e^A\wedge e^B
    =-\frac{1}{2}c^\mu_{\nu\lambda}e^\nu\wedge e^\lambda-c^\mu_{\nu n}e^\nu\wedge e^n
    =d_{\sss \Sigma}e^\mu-h e^\mu\wedge e^n,
\end{align*}
using the fact that $c^\mu_{\nu n} = h\delta^\mu_\nu$,
one has
\begin{align*} 
    d(e^{\mu_1}\wedge \cdots\wedge e^{\mu_t})
    &=\sum_{k=1}^t(-1)^{k-1}e^{\mu_1}\wedge \cdots\wedge (de^{\mu_k})\wedge \cdots\wedge e^{\mu_t}\\
     &=\sum_{k=1}^t(-1)^{k-1}e^{\mu_1}\wedge \cdots\wedge (d_{\sss \Sigma}e^{\mu_k}-h e^{\mu_k}\wedge e^n)\wedge \cdots\wedge e^{\mu_t}\\
  &=d_{\sss \Sigma}(e^{\mu_1}\wedge \cdots\wedge e^{\mu_t})
 -h\sum_{k=1}^t(-1)^{t-1}e^{\mu_1}\wedge  \cdots\wedge e^{\mu_t}\wedge e^n\\
   &=d_{\sss \Sigma}(e^{\mu_1}\wedge \cdots\wedge e^{\mu_t})
 +(-1)^{t} ht\ e^{\mu_1}\wedge  \cdots\wedge e^{\mu_t}\wedge e^n.
\end{align*}
The second identity is obtained from the first using the fact that $d e^n = 0$.
\end{proof}

\begin{property}\label{Prop2}
For $\tau\in \Omega^t(\Sigma)$,  
\begin{align*}
    &m^*\ast_{n+1} \tau\wedge e^{n}=\eta^{nn}(-1)^{n-t}\ast_{\sss \Sigma}\tau,\\
    &m^*\ast_{n+1} \tau =0.
\end{align*}
\end{property}
\begin{proof}
    By linearity, it is sufficient to check the property for $\tau=e^{\mu_1}\wedge\cdots\wedge e^{\mu_t}$. 
    First,
    \begin{align*}
    *_{n+1} e^{\mu_1}\wedge\cdots\wedge e^{\mu_t}\wedge e^{n}
    &= \eta^{\mu_1\nu_1}\cdots\eta^{\mu_t\nu_t}\eta^{nn}
        \epsilon_{\nu_1\cdots\nu_t n\rho_1\cdots\rho_{n-t}}
        \frac{1}{(n-t)!}e^{\rho_1}\wedge\cdots\wedge e^{\rho_{n-t}}\\
    &=\frac{(-1)^{n-t}}{(n-t)!}\eta^{nn}
        \eta^{\mu_1\nu_1}\cdots\eta^{\mu_t\nu_t}
        \epsilon_{\nu_1\cdots\nu_t\rho_1\cdots\rho_{n-t}n}
        e^{\rho_1}\wedge\cdots\wedge e^{\rho_{n-t}}\\
    &=\frac{(-1)^{n-t}}{(n-t)!}\eta^{nn}
        \eta^{\mu_1\nu_1}\cdots\eta^{\mu_t\nu_t}
        \epsilon_{\nu_1\cdots\nu_t\rho_1\cdots\rho_{n-t}}
        e^{\rho_1}\wedge\cdots\wedge e^{\rho_{n-t}}\\
    &=\eta^{nn}(-1)^{n-t} *_{\sss\Sigma} e^{\mu_1}\wedge\cdots\wedge e^{\mu_t},
\end{align*}
and
\begin{align*}
    \ast_{n+1} e^{\mu_1}\wedge\cdots\wedge e^{\mu_t}
    &= \eta^{\mu_1\nu_1}\cdots \eta^{\mu_t\nu_t}\frac{1}{(n-t)!}
        \epsilon_{\nu_1\ldots\nu_t \rho_1\ldots \rho_{n-t} n}
        e^{\rho_1}\wedge\cdots\wedge e^{\rho_{n-t}}\wedge e^{n}\\
    &= (*_{\sss\Sigma} e^{\mu_1}\wedge\cdots\wedge e^{\mu_t})\wedge e^n.
\end{align*}
Thus applying $m^*$ on both equations ends the proof.
\end{proof}

The remaining properties derived in this section rely heavily on Prop.~\ref{Prop1} and \ref{Prop2}, to which we will not refer to explicitly in the sequel.

\begin{property}\label{prop3}
    $m^*\delta_{n+1} e^{\mu_1}\wedge\cdots\wedge e^{\mu_t}
        = \delta_{\sss\Sigma}  e^{\mu_1}\wedge\cdots\wedge e^{\mu_t}.$    
\end{property}
\begin{proof}
    First, notice that:
    \begin{align*}
        *_{n+1} d *_{n+1} e^{\mu_1}\wedge\cdots e^{\mu_t}
        & = *_{n+1} d (*_{\sss\Sigma} e^{\mu_1}\wedge\cdots\wedge e^{\mu_t})\wedge e^n\\
        &= *_{n+1}  (d_{\sss\Sigma}*_{\sss\Sigma} e^{\mu_1}\wedge\cdots\wedge e^{\mu_t})\wedge e^n\\
        &= (-1)^{n-(n-t+1)}\eta^{nn} *_{\sss\Sigma}d_{\sss\Sigma}*_{\sss\Sigma}
            e^{\mu_1}\wedge\cdots\wedge e^{\mu_t}\\
        &= (-1)^{t+1}\eta^{nn} *_{\sss\Sigma}d_{\sss\Sigma}*_{\sss\Sigma}
            e^{\mu_1}\wedge\cdots\wedge e^{\mu_t},
    \end{align*}
    using the fact that $de^n = 0$ and that $(\cdots e^n)\wedge e^n = 0$.
    Paying attention to the signs yields then
    \begin{align*}
        \delta_{n+1} e^{\mu_1}\wedge\cdots\wedge e^{\mu_t}
        &= (-1)^{(n+1)(t+1)+1} \sgn(\eta_{n+1}) *_{n+1} d *_{n+1} e^{\mu_1}\wedge\cdots\wedge e^{\mu_t}\\
        &= (-1)^{(n+1)(t+1)+1} \sgn(\eta_{n+1}) (-1)^{t+1}\eta^{nn}
            *_{\sss\Sigma}d_{\sss\Sigma}*_{\sss\Sigma} e^{\mu_1}\wedge\cdots\wedge e^{\mu_t}\\
        &= (-1)^{(n+1)(t+1)+1} \sgn(\eta_{n+1}) (-1)^{t+1}\eta^{nn} (-1)^{n(t+1)+1} \sgn(\eta_{\sss\Sigma})
            \delta_{\sss\Sigma} e^{\mu_1}\wedge\cdots\wedge e^{\mu_t}\\
        &= \eta^{nn}\sgn(\eta_{n+1})\sgn(\eta_{\sss\Sigma})
            \delta_{\sss\Sigma} e^{\mu_1}\wedge\cdots\wedge e^{\mu_t}\\
        &= \delta_{\sss\Sigma} e^{\mu_1}\wedge\cdots\wedge e^{\mu_t},
    \end{align*}
    using the fact that $\eta^{nn}\sgn(\eta_{n+1})\sgn(\eta_{\sss\Sigma})=1$.
\end{proof}
\begin{property}\label{prop4}
    $m^*\delta_{n+1} e^{\mu_1}\wedge\cdots\wedge e^{\mu_t}\wedge e^n 
    = (-1)^{t+1}\eta^{nn}H(n-t)e^{\mu_1}\wedge\cdots\wedge e^{\mu_t}.$
\end{property}
\begin{proof}
    Notice that
    \begin{align*}
        &*_{n+1}d*_{n+1} e^{\mu_1}\wedge\cdots\wedge e^{\mu_t}\wedge e^n
         = (-1)^{n-t}\eta^{nn}*_{n+1}d (*_{\sss\Sigma} e^{\mu_1}\wedge\cdots\wedge e^{\mu_t})\\
        &\quad = (-1)^{n-t}\eta^{nn}*_{n+1}\bigl[
            (d_{\sss\Sigma}*_{\sss\Sigma} e^{\mu_1}\wedge\cdots\wedge e^{\mu_t})
        + (-1)^{n-t}h(n-t) (*_{\sss\Sigma} e^{\mu_1}\wedge\cdots\wedge e^{\mu_t})\wedge e^n \bigr]\\
        &\quad =  (-1)^{n-t}\eta^{nn}\bigl[
            (*_{\sss\Sigma}d_{\sss\Sigma}*_{\sss\Sigma} e^{\mu_1}\wedge\cdots\wedge e^{\mu_t})\wedge e^n
        + (-1)^{n-t}h(n-t) (-1)^{n-(n-t)}\eta^{nn}(*_{\sss\Sigma}*_{\sss\Sigma}
            e^{\mu_1}\wedge\cdots\wedge e^{\mu_t})\bigr]\\
        &\quad =  (-1)^{n-t}\eta^{nn}
            (*_{\sss\Sigma}d_{\sss\Sigma}*_{\sss\Sigma} e^{\mu_1}\wedge\cdots\wedge e^{\mu_t})\wedge e^n
        + (-1)^{nt}\sgn(\eta_{\sss\Sigma})h(n-t) e^{\mu_1}\wedge\cdots\wedge e^{\mu_t},
    \end{align*}
    applying $m^*$ cancels the first term and with the appropriate signs to have a codifferential it yields
    \begin{align*}
        m^*\delta_{n+1} e^{\mu_1}\wedge\cdots\wedge e^{\mu_t}\wedge e^n
        &= (-1)^{(n+1)((t+1)+1)+1}\sgn(\eta_{n+1})(-1)^{nt}\sgn(\eta_{\sss\Sigma})H(n-t)
            e^{\mu_1}\wedge\cdots\wedge e^{\mu_t}\\
        &= (-1)^{t+1}\eta^{nn}H(n-t)e^{\mu_1}\wedge\cdots\wedge e^{\mu_t},
    \end{align*}
    since $m^*h= H$ and $\eta^{nn}\sgn(\eta_{n+1})\sgn(\eta_{\sss\Sigma}) = 1$.
\end{proof} 
\begin{remark}
    Without applying $m^*$ to the equality one gets the ambient formula
    \begin{equation*}
        \delta_{n+1} e^{\mu_1}\wedge\cdots\wedge e^{\mu_t}\wedge e^n
        = (\delta_{\sss\Sigma} e^{\mu_1}\wedge\cdots\wedge e^{\mu_t})\wedge e^n
        + (-1)^{t+1}\eta^{nn}h(n-t) e^{\mu_1}\wedge\cdots\wedge e^{\mu_t}.
    \end{equation*}
\end{remark}

\begin{property}\label{prop5}
    Let $\tau\in \Omega^t(\Sigma)$, one has:  $i_n\ast_{n+1}  (\tau\wedge e^{n})=0.$
\end{property}
\begin{proof}
    $i_n *_{n+1} \tau\wedge e^n
    = (-1)^{n-t}\eta^{nn}i_n*_{\sss\Sigma}\tau
    = 0.$
\end{proof}
\begin{property}\label{prop6}
    $i_n\delta_{n+1} e^{\mu_1}\wedge\cdots\wedge e^{\mu_t} = 0.$
\end{property}
\begin{proof}
    $i_n\delta_{n+1} e^{\mu_1}\wedge\cdots\wedge e^{\mu_t}
    = i_n \delta_{\sss\Sigma} e^{\mu_1}\wedge\cdots\wedge e^{\mu_t}
    = 0.$
\end{proof}

\begin{property}\label{prop:Lie_n}
    Noting $\Lie_n = \Lie_{e_n} = i_nd+di_n$, the Lie derivative along $e_n$, one has
    \begin{align*}
        &\Lie_n e^{\mu_1}\wedge\cdots\wedge e^{\mu_t} 
        = ht\, e^{\mu_1}\wedge\cdots\wedge e^{\mu_t},\\
        &\Lie_n e^{\mu_1}\wedge\cdots\wedge e^{\mu_t}\wedge e^n
        = ht\, e^{\mu_1}\wedge\cdots\wedge e^{\mu_t}\wedge e^n.
    \end{align*}
\end{property}
\begin{proof}
    \begin{align*}
        \Lie_n e^{\mu_1}\wedge\cdots\wedge e^{\mu_t}
        = i_n de^{\mu_1}\wedge\cdots\wedge e^{\mu_t}
        = (-1)^tht i_n e^{\mu_1}\wedge\cdots\wedge e^{\mu_t}\wedge e^n
        = ht e^{\mu_1}\wedge\cdots\wedge e^{\mu_t},
    \end{align*}
    and since $d e^n = 0$ and then $\Lie_n e^n = di_ne^n = 0$, one has
    \begin{align*}
        \Lie_n e^{\mu_1}\wedge\cdots\wedge e^{\mu_t}\wedge e^n
        &= (\Lie_n e^{\mu_1}\wedge\cdots\wedge e^{\mu_t})\wedge e^n
        + e^{\mu_1}\wedge\cdots\wedge e^{\mu_t}\wedge (\Lie_n e^n)\\
        &= ht e^{\mu_1}\wedge\cdots\wedge e^{\mu_t}\wedge e^n.
        \qedhere
    \end{align*}
\end{proof}

\begin{property}\label{prop:Lrcorner}
    Let $\phi$ be a function, then
    \begin{align*}
        &d\phi \Lrcorner_{n+1} e^{\mu_1}\wedge\cdots\wedge e^{\mu_t}
        =  d_{\sss\Sigma}\phi \Lrcorner_{\sss\Sigma} e^{\mu_1}\wedge\cdots\wedge e^{\mu_t} \\
        &d\phi \Lrcorner_{n+1} e^{\mu_1}\wedge\cdots\wedge e^{\mu_t}\wedge e^n
        = (d_{\sss\Sigma}\phi\Lrcorner_{\sss\Sigma}e^{\mu_1}\wedge\cdots\wedge e^{\mu_t})\wedge e^n
        + (-1)^t \eta^{nn} e_n(\phi) e^{\mu_1}\wedge\cdots\wedge e^{\mu_t}.
    \end{align*}
\end{property}
\begin{proof}
    First, notice that 
    \begin{equation*}
        d\phi = d_{\sss\Sigma}\phi + e_n(\phi) e^n,
    \end{equation*}
    then 
    \begin{align*}
        d\phi\Lrcorner_{n+1} e^{\mu_1}\wedge\cdots\wedge e^{\mu_t}
        &= (-1)^{(n+1)(t+1)}\sgn(\eta_{n+1}) *_{n+1} d\phi\wedge *_{n+1} e^{\mu_1}\wedge\cdots\wedge e^{\mu_t}\\
        &= (-1)^{(n+1)(t+1)}\sgn(\eta_{n+1}) *_{n+1} d_{\sss\Sigma}\phi\wedge (*_{\sss\Sigma} e^{\mu_1}\wedge\cdots\wedge e^{\mu_t})\wedge e^n\\
        &= (-1)^{(n+1)(t+1)}\sgn(\eta_{n+1}) (-1)^{n-(1+n-t)}\eta^{nn}
             *_{\sss\Sigma} d_{\sss\Sigma}\phi\wedge (*_{\sss\Sigma} e^{\mu_1}\wedge\cdots\wedge e^{\mu_t})\\
        &= (-1)^{n(t+1)}\eta^{nn}\sgn(\eta_{n+1})(-1)^{n(t+1)}\sgn(\eta_{\sss\Sigma})
            d_{\sss\Sigma}\phi \Lrcorner_{\sss\Sigma} e^{\mu_1}\wedge\cdots\wedge e^{\mu_t}\\
        &= d_{\sss\Sigma}\phi \Lrcorner_{\sss\Sigma} e^{\mu_1}\wedge\cdots\wedge e^{\mu_t}.
    \end{align*}
    The second identity relies on the same relations and on a proper bookkeeping of signs.
\end{proof}

\section{Proofs of the theorems}\label{SEC-Proofs}
\subsection{Proof of Eq.~\eqref{EQ-Final-Restrict-delta}, restriction of $\delta_{n+1}$}

\begin{proof}
A key formula for the following proof is that for $\phi$ a function and $\beta\in\Omega^b$ a $b$-form one has
\begin{equation*}
    \delta(\phi\beta) = \phi (\delta\beta) - d\phi\Lrcorner\beta.
\end{equation*}
Then, owing to Prop.~\ref{prop4} and~\ref{prop:Lrcorner}, for a generic longitudinal term one has
\begin{align*}
    m^*\delta_{n+1}(\phi e^{\mu_1}\wedge\cdots\wedge e^{\mu_t}\wedge e^n)
    &= m^*\left[
        \phi\delta_{n+1}(e^{\mu_1}\wedge\cdots\wedge e^{\mu_t}\wedge e^n)
        - d\phi\Lrcorner_{n+1} e^{\mu_1}\wedge\cdots\wedge e^{\mu_t}\wedge e^n
    \right]\\
    &= m^*\left[
        \phi (-1)^{t+1}\eta^{nn}H(n-t)
        - (-1)^t \eta^{nn} e_n(\phi)   
    \right] e^{\mu_1}\wedge\cdots\wedge e^{\mu_t}\\
    &= (-1)^{t+1}\eta^{nn}m^*[e_n(\phi) + H(n-t)\phi]e^{\mu_1}\wedge\cdots\wedge e^{\mu_t}\\
    &= -\eta^{nn}m^*i_n[\Lie_n + H(n-2t)](\phi e^{\mu_1}\wedge\cdots\wedge e^{\mu_t}\wedge e^n),
\end{align*}
tweaking the last expression in order to recover the original longitudinal term in the right hand side, noting in particular with Prop.~\ref{prop:Lie_n} that
\begin{align*}
    [e_n(\phi) + h(n-t)\phi]e^{\mu_1}\wedge\cdots\wedge e^{\mu_t}
    &= (-1)^ti_n [e_n(\phi) + h(n-t)\phi] e^{\mu_1}\wedge\cdots\wedge e^{\mu_t}\wedge e^n\\
    &= (-1)^ti_n[\Lie_n +h(n-2t)](\phi e^{\mu_1}\wedge\cdots\wedge e^{\mu_t}\wedge e^n).
\end{align*}
For a generic transverse term
\begin{align*}
    m^*\delta_{n+1}(\phi e^{\mu_1}\wedge\cdots\wedge e^{\mu_t})
    &= m^*\left[
        \phi\delta_{n+1}(e^{\mu_1}\wedge\cdots\wedge e^{\mu_t})
        - d\phi\Lrcorner_{n+1} e^{\mu_1}\wedge\cdots\wedge e^{\mu_t}
    \right]\\
    &= m^*\left[
        \phi\delta_{\sss\Sigma}(e^{\mu_1}\wedge\cdots\wedge e^{\mu_t})
        - d_{\sss\Sigma}\phi\Lrcorner_{\sss\Sigma} e^{\mu_1}\wedge\cdots\wedge e^{\mu_t}
    \right]\\
    &= \delta_{\sss\Sigma}(m^*\phi e^{\mu_1}\wedge\cdots\wedge e^{\mu_t}).
\end{align*}
Thus, by linearity and the splitting of $\alpha\in\Omega^a(\setR^{n+1})$ in a longitudinal ($t=a-1$) and a transverse part ($t=a$), according to Eq.~\eqref{EQ-definition-AdaptedSpliting}, one gets
\begin{equation*}
    m^*\delta_{n+1}\alpha
    = \delta_{\sss\Sigma}m^*\alpha
    - \eta^{nn}m^*i_n[\Lie_n +H(n-2a-2)]\alpha,
\end{equation*}
recalling that $i_n\alpha_\perp = 0$, $m^*\alpha = \alpha_\perp = m^*\alpha_\perp$.
\end{proof}

\subsection{Proof of Eq.~\eqref{EQ-Final-Restrict-Box}, restriction of $\square_{n+1}$}

\begin{proof}
    Recalling that $m^*d\alpha = d_{\sss\Sigma}m^*\alpha$ [see Eq.~(6.2.12) of Ref.~\onlinecite{Fecko}],
    then one has
    \begin{align*}
        -m^*\square_{n+1}\alpha
        &= m^*(d\delta_{n+1} + \delta_{n+1}d)\alpha\\
        &= d_{\sss\Sigma}m^*\delta_{n+1}\alpha
        + \delta_{\sss\Sigma}d_{\sss\Sigma}m^*\alpha
        - \eta^{nn}m^*[\Lie_n + h(n-2a)]i_nd\alpha\\
        &= d_{\sss\Sigma}\delta_{\sss\Sigma}m^*\alpha
        - \eta^{nn}d_{\sss\Sigma}m^*[\Lie_n + h(n-2(a-1))]i_n\alpha
        + \delta_{\sss\Sigma}d_{\sss\Sigma}m^*\alpha
        - \eta^{nn}m^*[\Lie_n + h(n-2a)]i_nd\alpha\\
        &= -\square_{\sss\Sigma}\alpha_{\sss\Sigma}
        -\eta^{nn}m^*[(\Lie_n)^2 
            + h(n-2(a-1))\Lie_n
            - 2hi_nd
            ]\alpha
    \end{align*}
    using the fact that $d\Lie_n = \Lie_n d$, that $di_n = \Lie_n - i_nd$ and that $dh = -h^2 e^n$,
    this establishes the formula 
    \begin{align*}
        m^*\square_{n+1}\alpha
        &= \square_{\sss\Sigma}\alpha_{\sss\Sigma}
        +\eta^{nn}m^*[(\Lie_n)^2 
            + H(n-2(a-1))\Lie_n
            - 2Hi_nd
            ]\alpha\\
        &= \square_{\sss\Sigma}\alpha_{\sss\Sigma}
        +\eta^{nn}m^*[(\Lie_n)^2 
            + H(n-2a)\Lie_n
            + 2Hdi_n
            ]\alpha.
        \qedhere
    \end{align*}
\end{proof}
\begin{remark}
    Noting that since $e_n = h D$ and
    \begin{equation*}
        \Lie_n\alpha 
        = \Lie_{h \sss D}\alpha
        = h \Lie_{\sss D} + dh\wedge i_{\sss D}\alpha,
    \end{equation*}
    then $m^*\Lie_n\alpha = m^*H\Lie_{\sss D}\alpha$ and
    \begin{equation*}
        m^*(\Lie_n)^2\alpha
        = m^*h\Lie_{\sss D} h\Lie_{\sss D}\alpha
        =m^*H^2[(\Lie_{\sss D})^2 - \Lie_{\sss D}]\alpha
    \end{equation*}
    putting Eq.~\eqref{EQ-Final-Restrict-Box} as
    \begin{equation*}
        m^*\square_{n+1}\alpha
        = \square_{\sss\Sigma}\alpha_{\sss\Sigma}
        +\eta^{nn}H^2 m^*[(\Lie_{\sss D})^2 
            + (n - 1 - 2a)\Lie_{\sss D}
            + 2di_{\sss D}
            ]\alpha,
    \end{equation*}
    which for $a=1$ agrees with Eq.~(6) of Ref.~\onlinecite{Huguet:2022rxi}.
\end{remark}

\subsection{Proof of Eq.~\eqref{EQ-Final-Continuat-delta}, continuation of $\delta_{\sss \Sigma}$}

\begin{proof}
Let $\beta \in \Omega^b(\Sigma)$ and $\beta_s \in \Omega^b_{s\perp }$ such that 
$m^*\beta_s = \beta$,  its  $s$-homogeneous extension.
First, the restriction of the codifferential Eq.~\eqref{EQ-Final-Restrict-delta} together with 
the transversality of $\beta_s$ implies Eq.~\eqref{EQ-Final-Restrict-delta}, that is
\begin{equation*}
    m^*\delta_{n+1}\beta_s=   \delta_{\sss \Sigma}\beta.
\end{equation*}
Now, let us compute  $i_n\delta_{n+1}\beta_s$, one has:
\begin{align*}
   i_n\delta_{n+1}\beta_s
   &=\frac{1}{b!}i_n\delta_{n+1}(\beta_s)_{\mu_1\cdots\mu_b}e^{\mu_1}\wedge\cdots\wedge e^{\mu_b}\\
   &=\frac{1}{b!}[(\beta_s)_{\mu_1\cdots\mu_b}i_n\delta_{n+1}(e^{\mu_1}\wedge\cdots\wedge e^{\mu_b})
   -i_n\{(d(\beta_s)_{\mu_1\cdots\mu_b})\Lrcorner(e^{\mu_1}\wedge\cdots\wedge e^{\mu_b})\}]\\
  & =0,
\end{align*}
where, Prop.~\ref{prop6} and \ref{prop:Lrcorner} have been used in the last line.
Thus, the $(b-1)$-form
$\delta_{n+1}\beta_s$ is transverse (and since $\beta_s \in \Omega^b_{s\perp }$, 
$\delta_{n+1}\beta_s \in \Omega^{b-1}_{s-2 \perp }$),  which implies $m^*\delta_{n+1}\beta_s=\delta_{n+1}\beta_s$, and finally 
\begin{equation*}
     \delta_{\sss \Sigma}\beta =\delta_{n+1} \beta_s.
     \qedhere
\end{equation*}
\end{proof}

\subsection{Proof of Eq.~\eqref{EQ-Final-Continuat-Box}, continuation of $\square_{\sss \Sigma}$}

\begin{proof}
Let $\beta \in \Omega^b(\Sigma)$ and $\beta_s \in \Omega^b_{s\perp }$ such that 
$m^*\beta_s = \beta$,  its  $s$-homogeneous extension. 
Let us first compute $\L_n\beta_s$:
\begin{align*}
    \L_n\beta_s
    =\L_{\sss hD}\beta_s
    =h\L_{\sss D}\beta_s+dh\wedge i_{\sss D}\beta_s
    =hs\beta_s.
\end{align*}
Then, the restriction Eq.~\eqref{EQ-Final-Restrict-Box} becomes
\begin{align*}
    m^*\square_{n+1}\beta_s
    & =\square_{\sss \Sigma}\beta 
    + H^2\eta^{nn}[s^2+(n-1 -2b)s]\beta_s,\\
    & = \square_{\sss \Sigma}\beta 
    + H^2\eta^{nn}s(s + n -1 -2b)\beta_s.
\end{align*}
The longitudinal part of $\square_{n+1}\beta_s$ reads
\begin{align*}
    j^ni_n\square_{n+1}\beta_s
    &= -j^ni_n(d\delta_{n+1} + \delta_{n+1}d)\beta_s\\
    &= -j^n(\Lie_n\delta_{n+1} -di_n\delta_{n+1} -\delta_{n+1}i_nd)\beta_s\\
    &= -j^n(\Lie_n\delta_{n+1} -\delta_{n+1}\Lie_n)\beta_s\\
    &= -j^n(h(s-2) \delta_{n+1} - s\delta_{n+1}h)\beta_s\\
    &= -j^n(-2h\delta_{n+1}\beta_s + s dh\smash{\Lrcorner}\beta_s)\\
    &= 2h j^n\delta_{n+1}\beta_s\\
    &= 2h e^n\wedge (\delta_{n+1}\beta_s),
\end{align*}
in which we used $[i_n,d]_+=\L_n$, Prop.~\ref{prop6}, and
$dh=-h^2e^n$.

Now, since the pullback $m^*$, maps a $p$-form on $\setR^{n+1}$ to its transverse part
on $\Sigma$, on each point of $\Sigma$ one has
\begin{align*}
\square_{n+1}\beta_s&= 
(\square_{n+1}\beta_s)_\perp+ 
(\square_{n+1}\beta_s)_\parallel\\
&=m^*(\square_{n+1}\beta_s)+ j^ni_n \square_{n+1}\beta_s,
\end{align*}
from which Eq.~\eqref{EQ-Final-Continuat-Box} follows.
\end{proof}

%\bibliography{References}
%
\end{document}